\documentclass[aps,prb,groupedaddress,showpacs,showkeys]{revtex4} 
\usepackage{graphicx,natbib} 

\def\be{\begin{equation}} 
\def\ee{\end{equation}} 
\def\ba{\begin{eqnarray}} 
\def\ea{\end{eqnarray}} 
\def\bc{\begin{center}} 
\def\ec{\end{center}}

\begin{document} 

\title{Nonlinear electromagnetic response of graphene: \\ Frequency multiplication and the self-consistent-field effects} 

\author{S. A. Mikhailov} 
\email[Electronic mail: ]{sergey.mikhailov@physik.uni-augsburg.de} 
\affiliation{Institute for Physics, University of Augsburg, D-86135 Augsburg, Germany} 

\author{K. Ziegler} 
\affiliation{Institute for Physics, University of Augsburg, D-86135 Augsburg, Germany} 

\date{\today} 

\begin{abstract} 
Graphene is a recently discovered carbon based material with unique physical properties. This is a monolayer of graphite, and the two-dimensional electrons and holes in it are described by the effective Dirac equation with a vanishing effective mass. %graphene quasiparticles are massless Dirac fermions. 
As a consequence, electromagnetic response of graphene is predicted to be  strongly non-linear. We develop a quasi-classical kinetic theory of the non-linear electromagnetic response of graphene, taking into account the self-consistent-field effects. %and the radiative decay (electron-photon interaction). 
Response of the system to both harmonic and pulse excitation is considered. The frequency multiplication effect, resulting from the non-linearity of the electromagnetic response, is studied under realistic experimental conditions. The frequency up-conversion efficiency is analysed as a function of the applied electric field and parameters of the samples. Possible applications of graphene in terahertz electronics are discussed. 
\end{abstract} 


\maketitle 

\section{introduction} 

The region of the electromagnetic spectrum from 0.3 to 20 terahertz (so called Terahertz gap) is a frontier area for research in physics, chemistry, biology, material science and medicine\cite{opportunities}. Due to the recent progress in THz technology\cite{opportunities,Faist94,Siegel02}, the THz studies continue to expand, involving more and more scientists in the development of new sources and detectors of THz radiation, as well as in research on applications of THz waves in different areas. The search for new methods of THz emission and detection and development of simple, compact and inexpensive THz sources and detectors remains a challenging problem. 

The most common technique for producing low-power electromagnetic radiation at frequencies above 0.3 THz is through nonlinear multiplication (upconversion) of lower frequency oscillators \cite{Siegel02,Raisanen92}, $\Omega\to m\Omega$, $m=2,3,\dots$. Such upconverters, which are commonly based on GaAs Schottky barrier diodes, successfully work as doublers and triplers ($m=2$ and 3), but provide very poor conversion efficiency for higher order harmonics $(m>3)$ \cite{Siegel02,Raisanen92}. The search for alternative nonlinear materials, which could provide efficient frequency multiplication, especially with the upcoversion factor $m>3$, is therefore highly desirable. 

In this paper we discuss the recently predicted\cite{Mikhailov07e} effect of the frequency multiplication in graphene. Graphene is a new material, which has been experimentally obtained about three years ago\cite{Novoselov04,Novoselov05,Zhang05} and has immediately attracted great interest of researchers, for recent reviews see \cite{Katsnelson07,Geim07a}. Graphene is a monolayer of graphite, in which carbon atoms are packed in a dense two-dimensional (2D) honeycomb lattice\cite{Wallace47,Slonczewski58,Semenoff84}. This is a two-dimensional semimetal, with a very specific electronic band structure, Fig. \ref{bandstr}. The lower ($l=1$) and upper ($l=2$) branches of the energy spectrum $E_{{\bf p}l}$ touch each other at six points at the corners of the hexagon-shaped Brillouin zone. In the ideal case of a uniform, undoped graphene at zero temperature, the lower band $E_{{\bf p}1}$ is fully occupied while the upper band $E_{{\bf p}2}$ is empty, and the Fermi level goes through these six, so called Dirac points ${\bf P}^{(i)}=\hbar {\bf K}^{(i)}$, $i=1,\dots,6$. Near the Dirac points, at ${\bf p}\approx \hbar {\bf K}^{(i)}$, the electrons in graphene have a linear, quasirelativistic dispersion with zero effective mass of quasiparticles, 
\begin{figure} 
\caption{\label{bandstr} The band structure of electrons in graphene (side view and top view). The two branches of the energy spectrum $E_{{\bf p}l}$, $l=1,2$, touch each other at six points at corners of the hexagonal Brillouin zone. The values on the axis are $k_xa$ and $k_ya$, where $a$ is the lattice constant of graphene.} 
\end{figure} 
\be 
E_{{\bf p}l}=(-1)^l  V|{\bf p}-\hbar {\bf K}^{(i)}|=(-1)^l  V\sqrt{\tilde p_x^2+\tilde p_y^2},\label{spectrum} 
\ee 
where $\tilde {\bf p}={\bf p}-\hbar  {\bf K}^{(i)}$. Only two of the six Dirac points in the Brillouin zone, usually referred to as $K$ and $K'$ points, are inequivalent\cite{Wallace47,Slonczewski58,Semenoff84}. In the phenomena, discussed here, the states of the $K$ and $K'$ cones give additive contribution to the ac graphene response, so that one can consider only the states in the vicinity of one Dirac cone, accounting for the second cone by introducing the valley degeneracy factor $g_v=2$. The velocity $V$ in (\ref{spectrum}) is about $10^8$ cm/s, according to measurements\cite{Novoselov05,Zhang05}. Near the ${\bf K}^{(i)}$ points the graphene quasiparticles are described by two-component spinor wave functions, determined by the effective Dirac equation, and are called Dirac fermions.  

The unusual, linear dispersion of graphene quasiparticles near the Fermi level leads to a number of interesting and so far not fully understood transport phenomena, such as the minimal electrical conductivity\cite{Novoselov05,Zhang05,Katsnelson06,Peres06,Tworzydlo06,Nomura06,Ostrovsky06,Ziegler06,Cserti07,Ryu07,Nomura07,Ziegler07,Falkovsky07a,Cheianov07}, absense of the weak localization\cite{Morozov06}, unconventional quantum Hall effect\cite{Novoselov05,Zhang05,Gusynin05,Peres06,Gusynin06a,Gusynin07a}, observable even at room temperature\cite{Novoselov07}, and other. Electrodynamic properties of graphene, which have been studied both experimentally\cite{Sadowski06,Sadowski07,Bostwick07,Deacon07} and theoretically\cite{Falkovsky07a,Gusynin06a,Nilsson06,Peres06,Trauzettel07,Tworzydlo06,Gusynin06b,Gusynin07,Gusynin07a,Gusynin07b,Falkovsky07b,Abergel07,Hwang07,Wunsch06,Apalkov07,Ryzhii06,Ryzhii07,Vafek06,Mikhailov07d,Mikhailov07e}, also demonstrate non-trivial features in the 
frequency dependent conductivity\cite{Falkovsky07a,Gusynin06a,Nilsson06,Peres06,Mikhailov07d}, photon-assisted transport\cite{Trauzettel07}, microwave and far-infrared response\cite{Gusynin06b,Gusynin07,Gusynin07a,Gusynin07b,Falkovsky07b,Abergel07}, plasmons spectrum\cite{Hwang07,Wunsch06,Apalkov07,Ryzhii06,Ryzhii07}, et cetera. New electromagnetic modes, specific only for the graphene system, have been also predicted\cite{Vafek06,Mikhailov07d}. 

Both the transport and electrodynamic properties, briefly outlined above, have been studied within the linear response theory. Going beyond the linear-response approach, one can show\cite{Mikhailov07e,Mikhailov07f} that graphene should demonstrate {\em strongly nonlinear} electromagnetic response at relatively low amplitudes of the external electric field. In particular, irradiation of graphene by an electromagnetic wave with the frequency $\Omega$ should lead to the frequency multiplication $\Omega\to m\Omega$ with odd values of $m=3,5,7,\dots$. This makes graphene a simple and natural frequency multiplier\cite{Mikhailov07e,Mikhailov07f} and opens up exciting opportunities for using graphene in terahertz electronics.

In general, prospectives for building graphene based devices for terahertz applications are very attractive. Apart from the frequency multiplication, graphene devices could be used in {\em plasmon}-based voltage-controlled sources and detectors of THz radiation. The physics of such devices has long been discussed in the literature\cite{Dyakonov93,Dyakonov01,Shur03,Mikhailov98c,Kukushkin04a,Mikhailov-SPIE04} and the main ideas of 2D plasmon-assisted detection and generation of radiation have been confirmed in a number of experiments on conventional 2D electron systems\cite{Tsui80,Hirakawa95,Knap02b,Kukushkin05a}. Graphene has that advantage that the Fermi velocity of electrons in it is much higher than in other semiconductor materials, and its plasma frequency is widely tunable and lies in the THz range\cite{Ryzhii06,Ryzhii07}. It should be noticed that a closely related to graphene material - carbon nanotubes -- has also demonstrated a great potential for terahertz applications\cite{Slepyan99,Stanciu02,Kibis05a,Kibis05b,Nemilentsau06,Kuzhir07}. 

In this paper we discuss the effects of the non-linear electromagnetic response and the frequency multiplication in graphene\cite{Mikhailov07e,Mikhailov07f}. We consider response of graphene to a strong uniform external time-dependent electric field within the semi-classical kinetic theory (Section \ref{freqdom}). In addition to results obtained in Refs. \cite{Mikhailov07e,Mikhailov07f}, we take into account the self-consistent-field effects (Section \ref{self}), which should be important under realistic experimental conditions, and discuss both harmonic (Section \ref{freqdom}) and pulse excitation (Section \ref{timedom}) of the system. The results obtained are summarized and discussed in Section \ref{concl}.

\section{Self-consistent kinetic theory of non-linear electromagnetic response of graphene} 
\label{freqdom}

\subsection{Qualitative consideration} 

\begin{figure} 
\includegraphics[width=8.5cm]{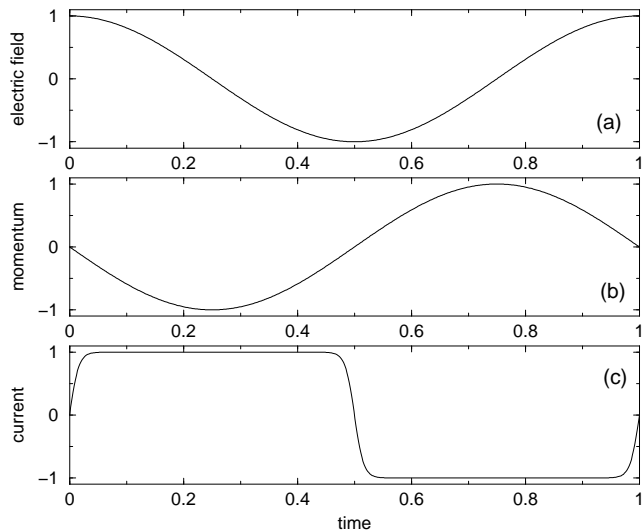} 
\caption{\label{sketch} Qualitative behavior of time dependencies of (a) the electric field, (b) the momentum, and (c) the velocity and current of a particle with the energy dispersion (\ref{ep}). } 
\end{figure}

Due to the linear dispersion (\ref{spectrum}) response of graphene to an external electromagnetic field turns out to be intrinsically non-linear, which naturally leads to the frequency multiplication effect. 
Physically, the possibility of the frequency upconversion in graphene can be explained in a very simple manner. Consider a classical 2D particle with the charge $-e$ and the energy spectrum
\be E_{{\bf p}2}=V p=V\sqrt{p_x^2+p_y^2}\label{ep}\ee  
in the external time-dependent harmonic electric field $E_x(t)=E_0\cos\Omega t$, Figure \ref{sketch}a. From now on we will consider only electrons in the vicinity of one Dirac cone, omit tilde in (\ref{spectrum}) and take into account the presence of two inequivalent cones in the Brillouin zone by the valley-degeneracy factor $g_v=2$. According to the Newton equation of motion $dp_x/dt=-eE_x(t)$ the momentum $p_x(t)$ is then given by the sine function 
\begin{equation} 
p_x(t)\equiv p_0(t)=-(eE_0/\Omega)\sin\Omega t, 
\end{equation} 
Figure \ref{sketch}b. In conventional 2D electron systems with the parabolic energy dispersion, the velocity $v_x$, and hence, the current $j_x=-en_sv_x$ are proportional to $p_x$, so that the normal 2D system responds at the same frequency $\Omega$ (here $n_s$ is the areal density of particles). In graphene, however, the velocity  
\begin{equation} 
v_x=\frac{\partial E_{{\bf p}2}}{\partial p_x}=V\frac{p_x}{\sqrt{p_x^2+p_y^2}}\label{vx} 
\end{equation} 
is a strongly non-linear function of $p_x$, therefore response of graphene is substantially anharmonic, Figure \ref{sketch}c. In the extreme limit, when $p_{y}$ in Eq. (\ref{vx}) is close to zero, $v_x$ is proportional to ${\rm sgn} (p_x)$ and the ac electric current is  
\begin{equation} 
j_x(t)= en_sV {\rm sgn} (\sin \Omega t)=en_sV \frac 4\pi\left\{\sin\Omega t + \frac 13 \sin 3\Omega t + \frac 15 \sin 5\Omega t + \dots\right\}.\label{estim} 
\end{equation} 
The current (\ref{estim}) contains all odd Fourier harmonics, with the amplitudes $j_m$, $m=1,3,5\dots$, falling down very slowly with the harmonics number, $j_m\sim 1/|m|$. An isolated  graphene layer should thus work as a simple and natural frequency multiplier, with the operating frequency, variable in a broad range.  

\subsection{Kinetic approach}\label{kinappr}

The current (\ref{estim}) above is independent of the electric field, which means that Eq. (\ref{estim}) is not completely correct. The reason is that in the above simple  consideration we did not take into account the Fermi distribution of electrons over the quantum states in graphene. To do this, we use\cite{Mikhailov07e} the kinetic Boltzmann theory, which allows one to get an exact response of the system not imposing any restrictions on the amplitude of the external electric field ${\bf E}^{ext}(t)$.

In a real experimental situation, graphene sheet lies on top of a silicon oxide --  silicon  structure, and the gate voltage $V_G$ can be applied between graphene and the silicon substrate, in order to control the density of electrons or holes in graphene. In addition, the graphene-SiO$_2$-Si system can be doped by impurities. Both the gate voltage and the doping can shift the chemical potential $\mu$ of electrons in graphene to the upper $E_{{\bf p}2}$ or to the lower $E_{{\bf p}1}$ band. Assume that the chemical potential $\mu$ lies in the upper band $E_{{\bf p}2}=Vp$, the temperature is small, $T\ll \mu$, and the system is subjected to the external time-dependent ac electric field ${\bf E}^{ext}(t)$. 
Then the momentum distribution function of electrons $f_{\bf p}(t)$ is described by the Boltzmann equation 
\begin{equation} 
\frac{\partial f_{\bf p}(t) }{\partial t} -\frac{\partial f_{\bf p}(t)}{\partial {\bf p}} e {\bf E}^{ext}(t)=0, %0 e^{\alpha t} \cos(\Omega t)=0, 
\label{be}
\end{equation} 
in which we have ignored collisions of electrons with impurities, phonons and other lattice imperfections. 
Equation (\ref{be}) has the exact solution 
\begin{equation} 
f_{\bf p}(t) = {\cal F}_0\left({\bf p-p}_0(t)\right), \label{foft}
\end{equation} 
where %${\cal F}_0({\bf p})$ is the Fermi-Dirac function
\begin{equation} 
{\cal F}_0({\bf p}) =\left[1+\exp\left(\frac{ Vp%\sqrt{p_x^2+p_y^2}
-\mu}T\right)\right]^{-1} 
\end{equation} 
is the Fermi-Dirac function, and 
\be 
{\bf p}_0(t)=-e\int^t_{-\infty} {\bf E}^{ext}(t')dt'
\ee 
is the solution of the single particle classical equation of motion. The electric current 
\be
{\bf j}(t)=-\frac{eg_sg_v}S \sum_{{\bf p}} {\bf v} f_{\bf p}(t)
\ee 
then assumes the form 
\begin{equation} 
{\bf j}(t)= -\frac{g_sg_veV}{(2\pi \hbar)^2}\int \frac{{\bf p}d{\bf p}}p {\cal F}_0\left({\bf p-p}_0(t)\right),
\label{jx1} 
\end{equation} 
where $g_s=2$ is the spin degeneracy and $S$ is the sample area. If the temperature is zero, $T=0$, and the chemical potential is finite, $\mu>0$, the current ${\bf j}(t)$ can be rewritten in the form  
\begin{equation} 
{\bf j}(t) = en_sV \frac{\bf P}{\sqrt{1+P^2}}{\cal G}(Q),
\label{jx2} 
\end{equation} 
where ${\bf P}\equiv {\bf P}(t)=-{\bf p}_0(t)/p_F$, $P(t)=|{\bf P}(t)|$, $p_F=\mu/V$  is the Fermi momentum, and 
\begin{equation} 
n_s\equiv n_e=\frac{g_sg_vp_F^2}{4\pi \hbar^2}=\frac{g_sg_v\mu^2}{4\pi \hbar^2 V^2} 
\end{equation} 
is the density of electrons in the upper band. The function ${\cal G}(Q)$ in (\ref{jx2}) is defined as
\be
{\cal G}(Q)=\frac 4{\pi Q}
\int_{0}^{\pi/2}  
\cos x dx (\sqrt{1+Q\cos x} - 
\sqrt{1-Q\cos x}), \ \ Q(t)\equiv \frac{2P(t)}{1+P^2(t)}\le 1,
\label{funG}
\ee 
and analyzed in Appendix \ref{funcG}. 

If the external field ${\bf E}^{ext}$ is small, so that the Fermi distribution is weakly disturbed, $P(t)\ll 1$, the function $G(Q)\approx 1$, see Eq. (\ref{G}), and 
\be
{\bf j}(t)\approx\frac{n_se^2V}{p_F}\int^t_{-\infty}{\bf E}^{ext}(t')dt'.
\ee
In this, {\em linear response} regime, the current is directly proportional to the electric field. If the external excitation is harmonic, 
\be 
{\bf E}^{ext}(t)={\bf E}_0\cos\Omega t, \label{harm}
\ee 
the current is
\be
{\bf j}(t)\approx\frac{n_se^2V}{p_F\Omega}{\bf E}_0\sin\Omega t,
\ee
which corresponds to the linear-response frequency-dependent collisionless scalar Drude conductivity\cite{Gusynin06a,Gusynin06b,Gusynin07,Gusynin07a,Falkovsky07a,Mikhailov07d}
\be
\sigma(\Omega)=\frac {in_se^2V}{\Omega p_F}=i\frac{e^2}\hbar\frac{g_sg_v}{4\pi}\frac{\mu}{\hbar\Omega}.\label{lincond} 
\ee

If the external field ${\bf E}^{ext}$ is large and the Fermi distribution is strongly disturbed, $P(t)$ can be much larger than 1, the function ${\cal G}(Q)\approx 1$, and response of the system to a harmonic excitation (\ref{harm}) is described by Eq. (\ref{estim}). The non-linear regime (\ref{estim}) is thus achieved at $|p_0|\gg p_F$, or at
\be
{\cal E}\equiv \frac{eE_0V}{\Omega\mu}\gg 1:
\label{strongfield}
\ee
the energy, gained by electrons from the external field during the oscillation period should be large as compared to their average equilibrium (Fermi) energy. 

\begin{figure} 
\includegraphics[width=8.5cm]{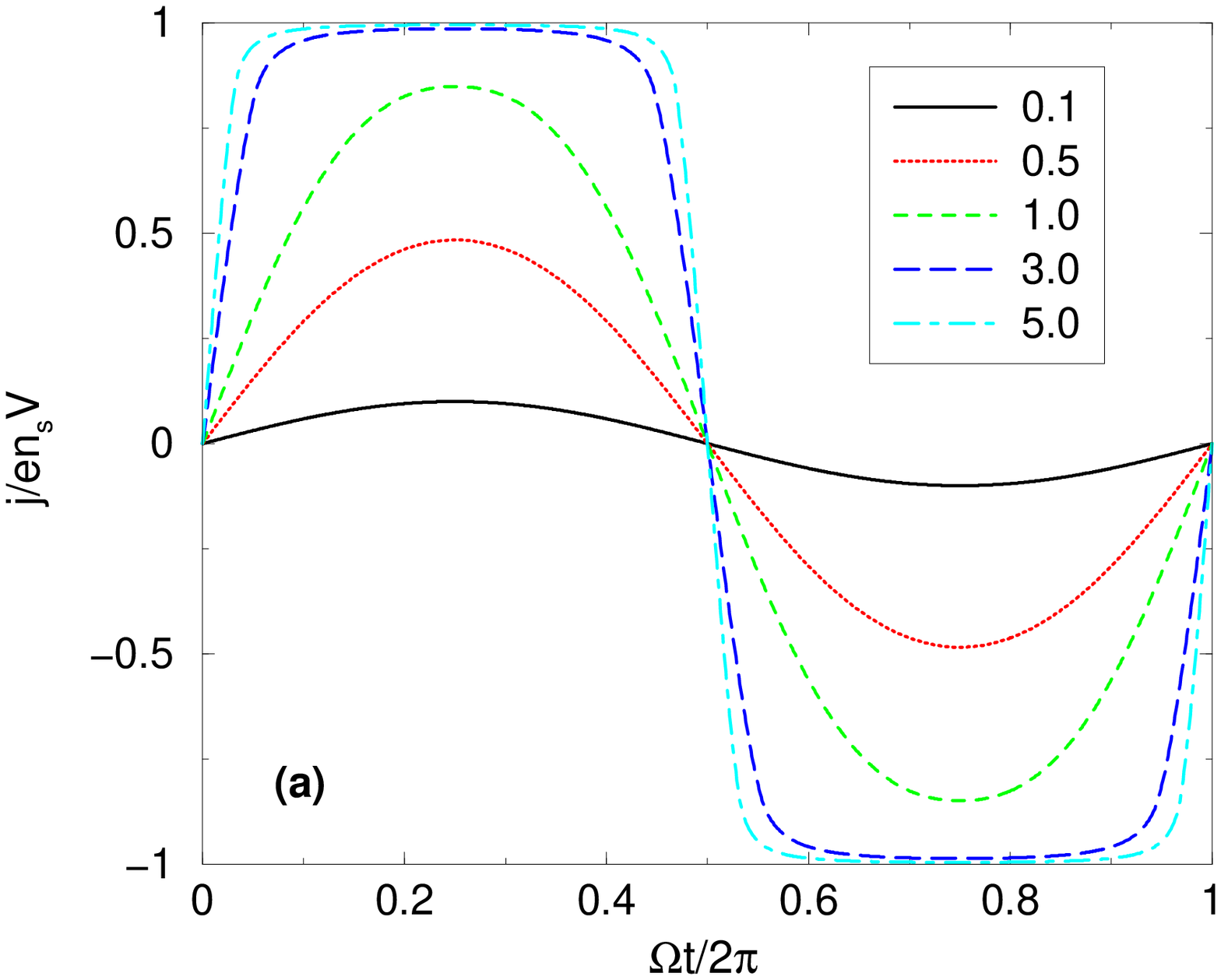}\includegraphics[width=8.5cm]{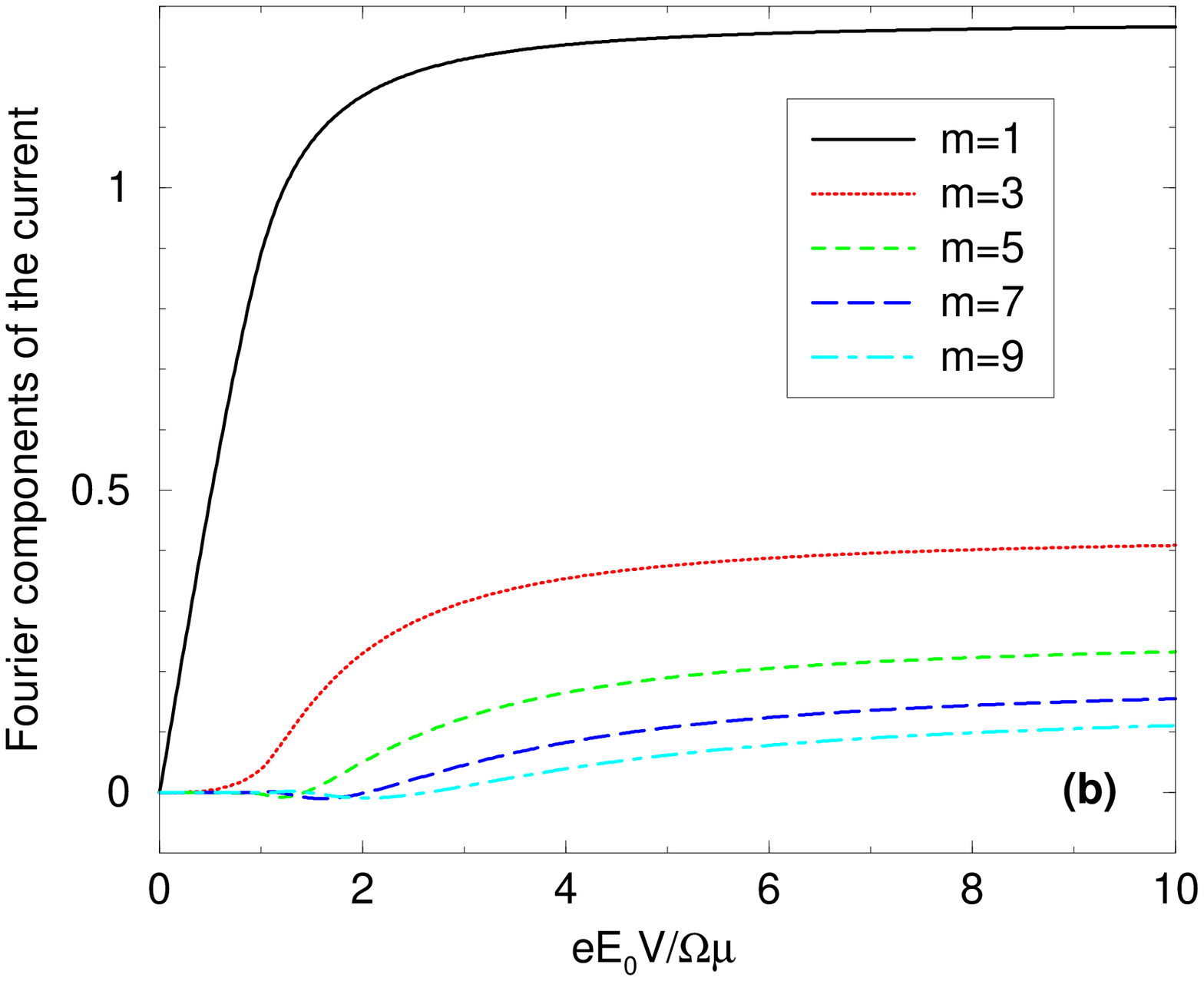} 
\caption{\label{T0} (Color online) (a) The time dependence of the ac electric current, measured in units $en_sV$, at harmonic excitation of the system at the frequency $\Omega$. The temperature is zero, $T/\mu=0$; the curves are labeled by the values of the electric field parameter ${\cal E}=eE_0V/\Omega \mu$. (b) The Fourier components of the current (\ref{jx2}) as a function of ${\cal E}$ at $T/\mu=0$. At ${\cal E}_T\to\infty$ the curves tend to the values $4/\pi m$, see Eq. (\ref{estim}).
} 
\end{figure} 

Figure \ref{T0}a illustrates the time dependence of the current (\ref{jx2}) at the harmonic excitation of the system (\ref{harm}). One sees that in the low-field limit the response is linear (the $j(t)$ dependence has a sinusoidal shape), while at strong fields the time dependence of the current tends to that given by Eq. (\ref{estim}). In Figure \ref{T0}b  we show the Fourier components of the current versus the field parameter ${\cal E}$. One sees that when ${\cal E}$ becomes larger than $\simeq 4$, the Fourier amplitudes saturate and one gets in the ultimate non-linear regime. 

The strong-field condition (\ref{strongfield}) can be rewritten as 
\begin{equation} 
E_0\gg
\frac{2 \hbar \Omega\sqrt{\pi n_s}}{e\sqrt{g_sg_v}}, 
\label{condition} 
\end{equation} 
which shows that the required ac electric field grows linearly with the electromagnetic wave frequency and with the square root of the electron density. At $f\simeq 100$ GHz and $n_s\simeq 10^{11}$ cm$^{-2}$, the inequality (\ref{condition}) is fulfilled at $E_0\gtrsim 200$ V/cm. 

The results obtained above refer to the case $T/\mu=0$ ($\mu$ is finite). The opposite limit $\mu=0$ (at a finite temperature, $\mu/T=0$) is difficult to realize in the whole sample, at least, in currently available systems. Due to inhomogeneity of the samples, the zero-energy point fluctuates in space, with a typical local electron or hole density of order of $10^{11}$ cm$^{-2}$, Ref. \cite{Martin07}. The condition $\mu/T=0$ can thus be satisfied only locally, at certain points inside the sample. Nevertheless, for the sake of completeness (and in view of possible availability of higher quality samples in future), we also present here results\cite{Mikhailov07e} for the case $\mu/T=0$. For arbitrary values of $\mu/T$ some results can be also found in Ref. \cite{Mikhailov07f}.

At a finite temperature and the vanishing chemical potential $\mu=0$ both electrons and holes contribute to the charge carrier density 
\begin{equation} 
n_s=n_e+n_h=\frac{\pi g_sg_vT^2}{12\hbar^2V^2} 
\end{equation} 
and to the current. Starting again from Eq. (\ref{jx1}) but accounting for the hole contribution and setting $\mu=0$, we get 
\begin{equation} 
{\bf j}(t)= 
en_sV \frac{{\bf P}_T}{P_T}\frac{12}{\pi^3}\int_0^\infty xdx \int_{0}^\pi d\theta  \frac {\cos\theta}{1+\exp\left(\sqrt{x^2+P_T^2-2x P_T\cos\theta}\right)},\label{jx3} 
\end{equation} 
where the momentum ${\bf p}_0(t)$ is now normalized by the characteristic temperature-dependent momentum, ${\bf P}_T\equiv {\bf P}_T(t)=-{\bf p}_0(t)/p_T$, $P_T(t)=|{\bf P}_T(t)|$, and $p_T=T/V$.
 
If the external field is small, $P_T(t)\ll 1$, Eq. (\ref{jx3}) gives the linear-response current
\begin{equation} 
{\bf j}(t)\approx en_sV {\bf P}_T(t)\frac {6\ln 2}{\pi^2}
=\frac{n_se^2V^2}{T}\frac {6\ln 2}{\pi^2}\int_{-\infty}^t{\bf E}^{ext}(t')dt'
, 
\end{equation} 
proportional to the electric field. If the external excitation is harmonic, Eq. (\ref{harm}), the current is
\be
{\bf j}(t)=\frac{n_se^2V^2}{T\Omega}\frac {6\ln 2}{\pi^2}{\bf E}_0\sin\Omega t
, 
\ee
which corresponds to the linear-response frequency-dependent collisionless Drude conductivity\cite{Falkovsky07a}
\begin{equation} 
\sigma_{\mu=0,T}(\Omega)=\frac {6\ln 2}{\pi^2}\frac{in_se^2V^2}{T\Omega} 
=i\frac {\ln 2}{2\pi}\frac{e^2}\hbar\frac{g_sg_vT}{\hbar\Omega}. 
\end{equation} 
In the strong field regime $P_{T0}\gtrsim 1$ Eq. (\ref{jx3}) is reduced, again, to (\ref{estim}). The dimensionless field parameter in the case $\mu/T=0$ is defined as ${\cal E}_T=eE_0V/T\Omega$.

\begin{figure} 
\includegraphics[width=8.5cm]{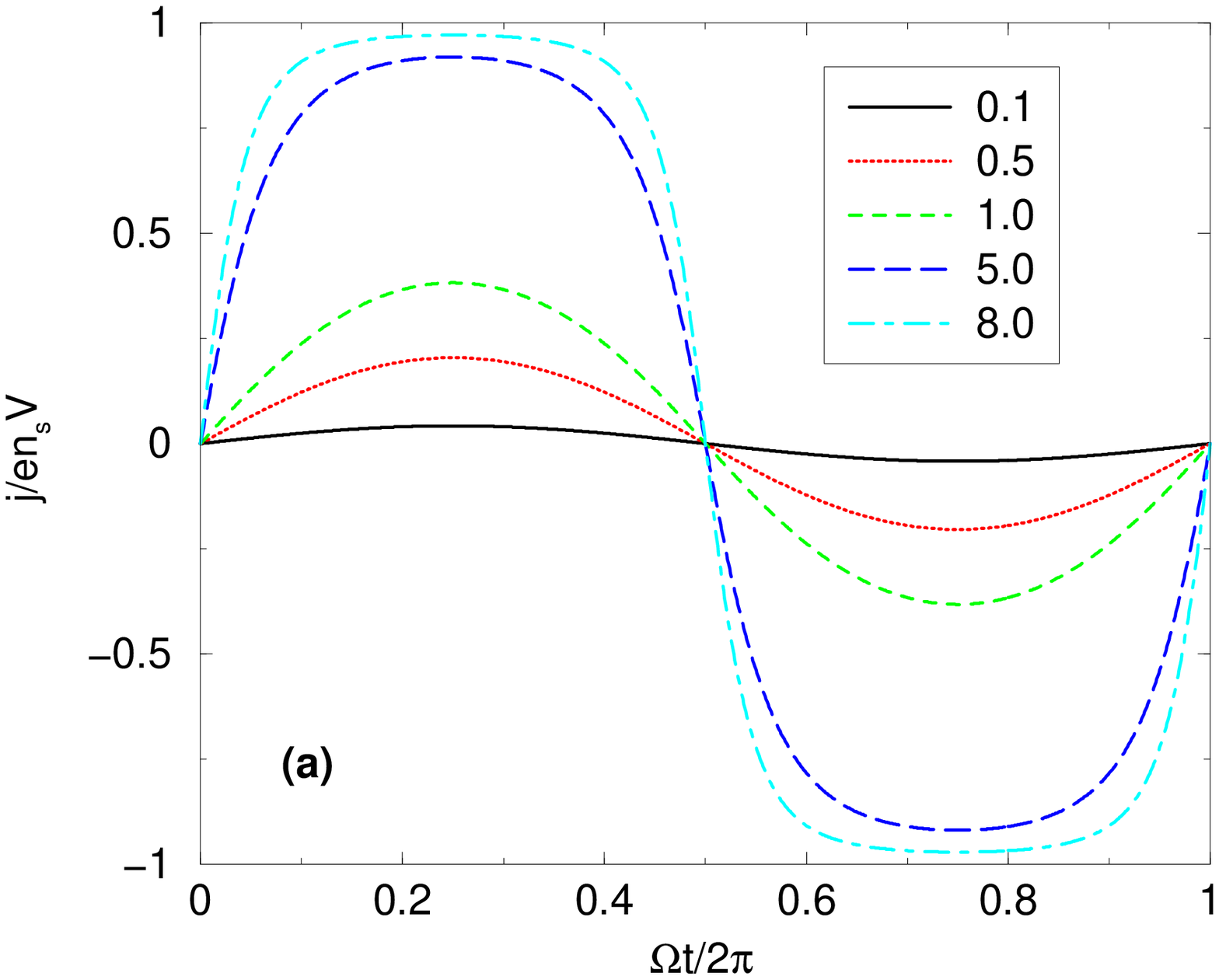}\includegraphics[width=8.5cm]{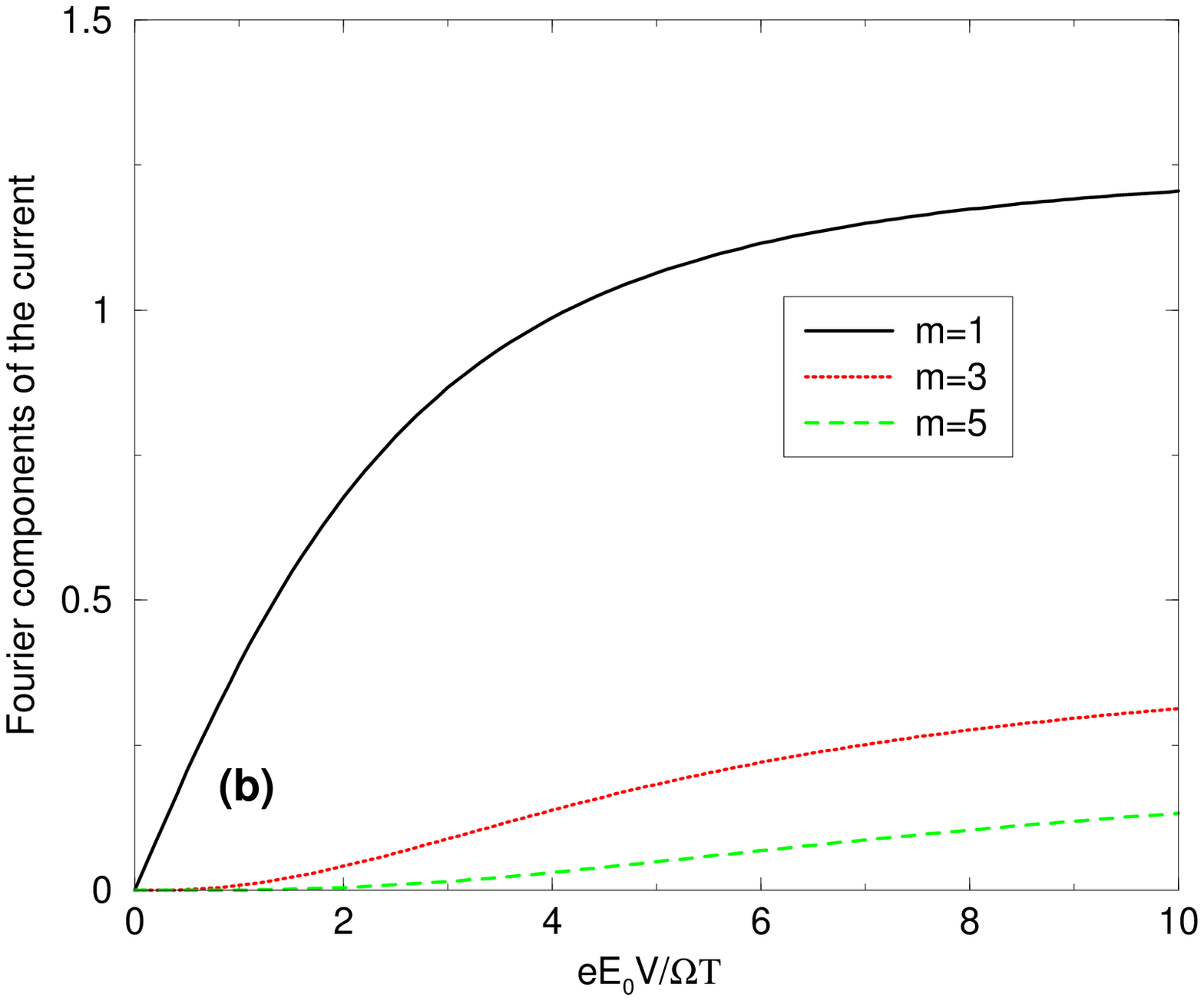}\\ 
\caption{\label{Mu0} (Color online) (a) The time dependence of the ac electric current, measured in units $en_sV$, at harmonic excitation of the system at the frequency $\Omega$. The chemical potential is zero and the temperature is finite, $\mu/T=0$; the curves are labeled by the values of the electric field parameter ${\cal E}_T=eE_0V/\Omega T$. (b) The Fourier components of the current (\ref{jx3}) as a function of the field parameter ${\cal E}_T=eE_0 V/\Omega T$ at $\mu/T=0$. At ${\cal E}_T\to\infty$ the curves tend to the values $4/\pi m$, see Eq. (\ref{estim}).} 
\end{figure}

Figure \ref{Mu0}a demonstrates the time dependence of the current (\ref{jx3}) at the harmonic excitation of the system (\ref{harm}). The $j(t)$ dependence is qualitatively similar to that, shown in Figure \ref{T0}a. In Figure \ref{Mu0}b we show the Fourier components of the ac electric current, for $m=1$, 3 and 5, as a function of the field parameter ${\cal E}_T$ at $\mu=0$.

In the case $\mu/T=0$ the strong-field condition ${\cal E}\gtrsim 1$ assumes the form 
\begin{equation} 
E_0\gtrsim\frac{\Omega T}{eV}. 
\end{equation} 
At the temperature $T\simeq 100$ K and the frequency $f\simeq 100$ GHz this corresponds to the requirement $E_0\gtrsim 50$ V/cm. 

Using the quasi-classical (Boltzmann) approach for the description of electromagnetic response of graphene imposes certain restrictions on the validity of the presented theory. Physically, using the Boltzmann theory, one takes into account the intra-band contribution to the ac electric current, but ignores the inter-band contribution due to transitions between the lower (quasi-hole) and the upper (quasi-electron) bands. This is possible if the frequency of the external radiation satisfies the inequality
\be 
\hbar\Omega\ll\max\{\mu,T\}.
\ee 
In practically interesting cases (the density $n_s\simeq 10^{11}-10^{12}$ cm$^{-2}$, room temperature) this restricts the frequency of radiation by $\sim 10-30$ THz. The presented quasi-classical theory can thus be used in the whole Terahertz gap.  

\subsection{Self-consistent-field effects and radiative decay}\label{self}

In Section \ref{kinappr} we have not considered effects of the radiative decay, which can be important under realistic experimental conditions in graphene. It was assumed that the graphene electrons move in the sample under the action of the external electric field ${\bf E}^{ext}(t)$, and this directly leads to the time-dependent electric current ${\bf j}(t)$, Eq. (\ref{jx2}). In general, however, the time-dependent electric current creates, in its turn, a secondary (induced) electric field ${\bf E}^{ind}(t)$, which acts back on the electrons and should be %selfconsistently 
added to the external field. Calculating response of the system, one should take into account that electrons respond not to the external, but to the total self-consistent electric field ${\bf E}%_{z=0}
^{tot}(t)={\bf E}^{ext}%_{z=0}
(t)+{\bf E}^{ind}%_{z=0}
(t)$. This results in the effects of electromagnetic reaction of the medium (graphene) to the external field and can reduce the frequency upconversion efficiency. How substantially suppresses the radiative decay the efficiency of the frequency multiplication, is studied below.

Consider an infinite 2D electron system with the graphene sheet lying at the plane $z=0$. We assume that the external electromagnetic wave is incident upon the graphene layer along the $z$ axis and induces the ac current in the layer. This current produces the induced electric field ${\bf E}^{ind}(t)$, which is added to the external one. The Boltzmann equation for the momentum distribution function of the electrons should then be written as
\begin{equation} 
\frac{\partial f_{\bf p}(t) }{\partial t} -\frac{\partial f_{\bf p}(t)}{\partial {\bf p}} e {\bf E}^{tot}_{z=0}(t)=0,\label{BE} 
\end{equation} 
instead of (\ref{be}). The solution of this equation, as well as the electric current, can again be written in the form (\ref{foft}) and (\ref{jx1}), respectively, but the classical momentum ${\bf p}_0(t)$ now satisfies the equation
\be 
{\bf p}_0(t)=-e\int^t_{-\infty} {\bf E}^{tot}_{z=0}(t')dt'=-e\int^t_{-\infty} \left[{\bf E}^{ind}_{z=0}(t')+{\bf E}^{ext}_{z=0}(t')\right]dt',\label{p0eq} 
\ee 
where the field ${\bf E}^{tot}_{z=0}(t)$ is not known and should be calculated self-consistently. To do this, we recall that the current and the induced electric field are related by the Maxwell equations, 
\be 
{\bf E}^{ind}_{z=0}(t)=-2\pi{\bf j}(t)/c. \label{EindJ} 
\ee 
Combining (\ref{p0eq}), (\ref{EindJ}) and (\ref{jx1}) we get the following self-consistent equation of motion for the momentum ${\bf p}_0(t)$: 
\be 
\frac{d{\bf p}_0(t)}{dt}+\frac{e^2g_sg_vV}{2\pi \hbar^2c}\int \frac{{\bf p}d{\bf p}}{p} {\cal F}_0\left({\bf p}-{\bf p}_0(t)\right)=-e{\bf E}^{ext}_{z=0}(t).\label{eqp0} 
\ee 
After the non-linear equation (\ref{eqp0}) is resolved with respect to the momentum ${\bf p}_0(t)$, the current ${\bf j}(t)$ can be found from Eq. (\ref{jx1}):
\begin{equation} 
{\bf j}(t)= -\frac{g_sg_veV}{(2\pi \hbar)^2}\int \frac{{\bf p}d{\bf p}}p {\cal F}_0\left({\bf p-p}_0(t)\right),
\label{j1} 
\end{equation} 
Equations (\ref{eqp0}) and (\ref{j1}) describe the non-linear self-consistent response of graphene to an arbitrary external time-dependent electric field ${\bf E}^{ext}_{z=0}(t)$.  The second term in the left-hand side of Eq. (\ref{eqp0}) describes the radiative decay effects in the infinite 2D graphene layer.

In conventional 2D electron systems with the parabolic energy dispersion and the effective mass $m^\star$ of 2D electrons the self-consistent equations for ${\bf p}_0(t)$ and ${\bf j}(t)$, similar to (\ref{eqp0}) and (\ref{j1}), have the form
\be 
\frac{d{\bf p}_0(t)}{dt}+%\Gamma_p 
\frac{2\pi n_se^2}{m^\star c}{\bf p}_0(t)=-e{\bf E}^{ext}_{z=0}(t),\ \ \ 
{\bf j}(t)=-\frac{en_s}{m^\star}{\bf p}_0(t). \label{conven}
\ee
Here 
$\Gamma_{par} \equiv 2\pi n_s e^2/m^\star c$ is the radiative decay rate \cite{Mikhailov04a} in the conventional ({\em par}abolic) 2D electron systems. In (\ref{conven}) we have ignored the scattering due to impurities and phonons [the corresponding term $\gamma{\bf p}_0(t)$ can be added to the left-hand-side of the first Eq. (\ref{conven})]. In high-electron mobility GaAs/AlGaAs quantum-well samples the radiative decay $\Gamma_{par}$ substantially exceeds the scattering rate $\gamma$, $\Gamma_{par}\gg\gamma$, and determines the linewidth of the cyclotron, plasmon, and magnetoplasmon resonances\cite{Mikhailov96a,Mikhailov04a}. As the graphene mobility is also very high, one can expect that at high frequencies the radiative effects are more important in graphene than the scattering effects. This justifies ignoring the scattering terms in Eq. (\ref{eqp0}).

Returning back to the non-parabolic graphene system, we rewrite (\ref{eqp0}) (at $T=0$) in terms of the dimensionless momentum ${\bf P}(t)= -{\bf p}_0(t)/p_F$:
\be 
\frac{d{\bf P}(t)}{dt}+\Gamma \frac{\bf P}{\sqrt{1+P^2}}{\cal G}(Q) 
=\frac{e}{p_F}{\bf E}^{ext}_{z=0}(t),\label{eqp0a} 
\ee 
where 
\be 
\Gamma=\frac{g_sg_v}4\frac{e^2}{\hbar c}\frac{2\mu}{\hbar}=V\frac{e^2}{\hbar c}\sqrt{g_sg_v\pi n_s}.\label{raddecgraph} 
\ee 
The current is determined, again, by Eq. (\ref{jx2}). 

In the linear-response regime, when $|P|\ll 1$ and ${\cal G}\approx 1$, Eq. (\ref{eqp0a}) gives 
\be 
\frac{d{\bf P}(t)}{dt}+\Gamma {\bf P}(t)=\frac e{p_F}{\bf E}^{ext}_{z=0}(t).\label{linear} 
\ee 
From here one sees that the quantity $\Gamma$ has the physical meaning of the radiative decay rate in graphene {\em in the linear-response regime}. In contrast to $\Gamma_{par}$, $\Gamma$ is proportional to the square-root of the charge carrier density. For experimentally relevant densities $n_s$ the value of $\Gamma$ lies in the subterahertz range, $\Gamma/2\pi$(THz)$\approx 0.13\sqrt{n_s (10^{11}/{\rm cm}^{2})}$. % at $n_s\simeq 10^{12}$ cm$^{-2}$). 

\begin{figure} 
\includegraphics[width=8.5cm]{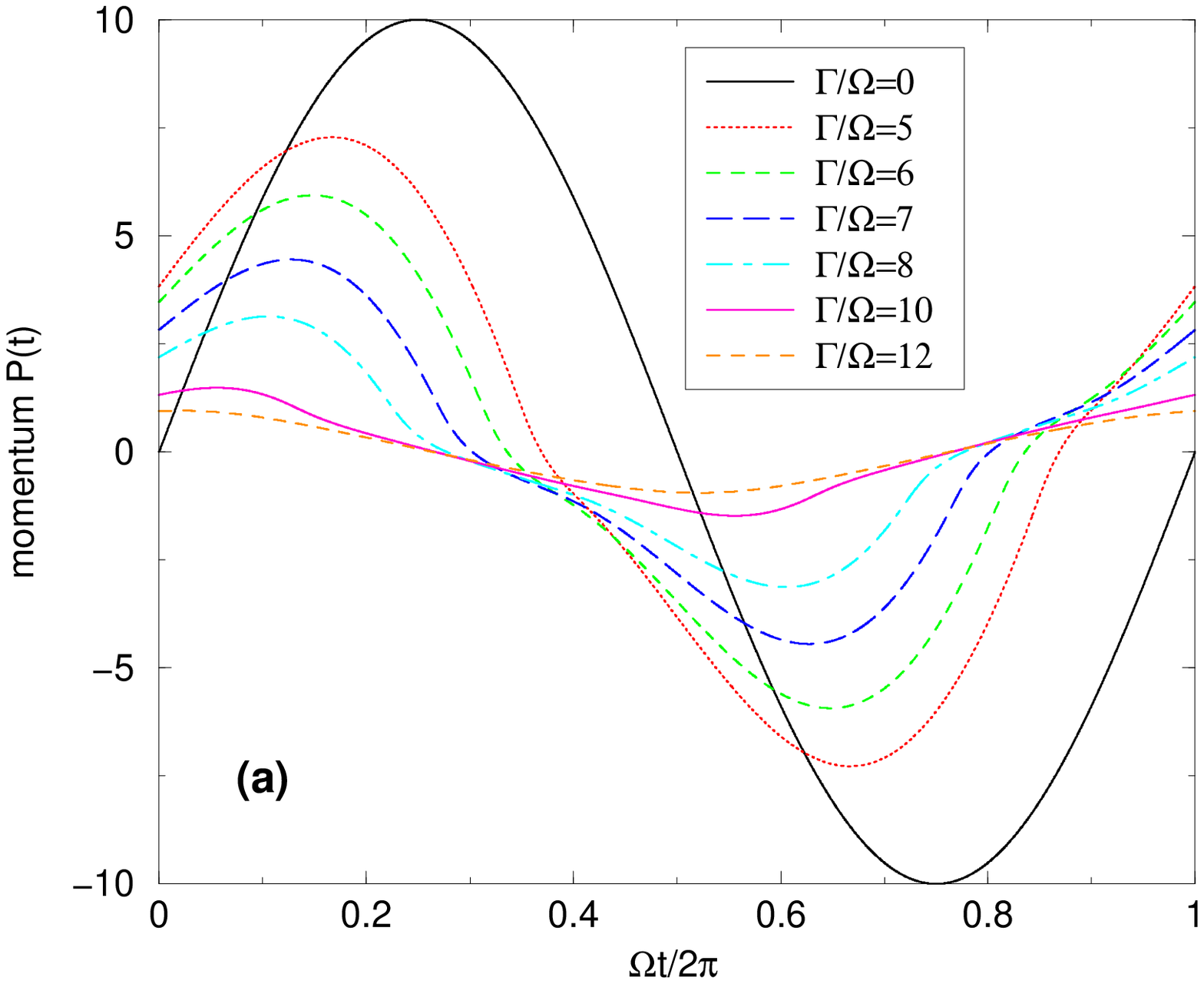} \includegraphics[width=8.5cm]{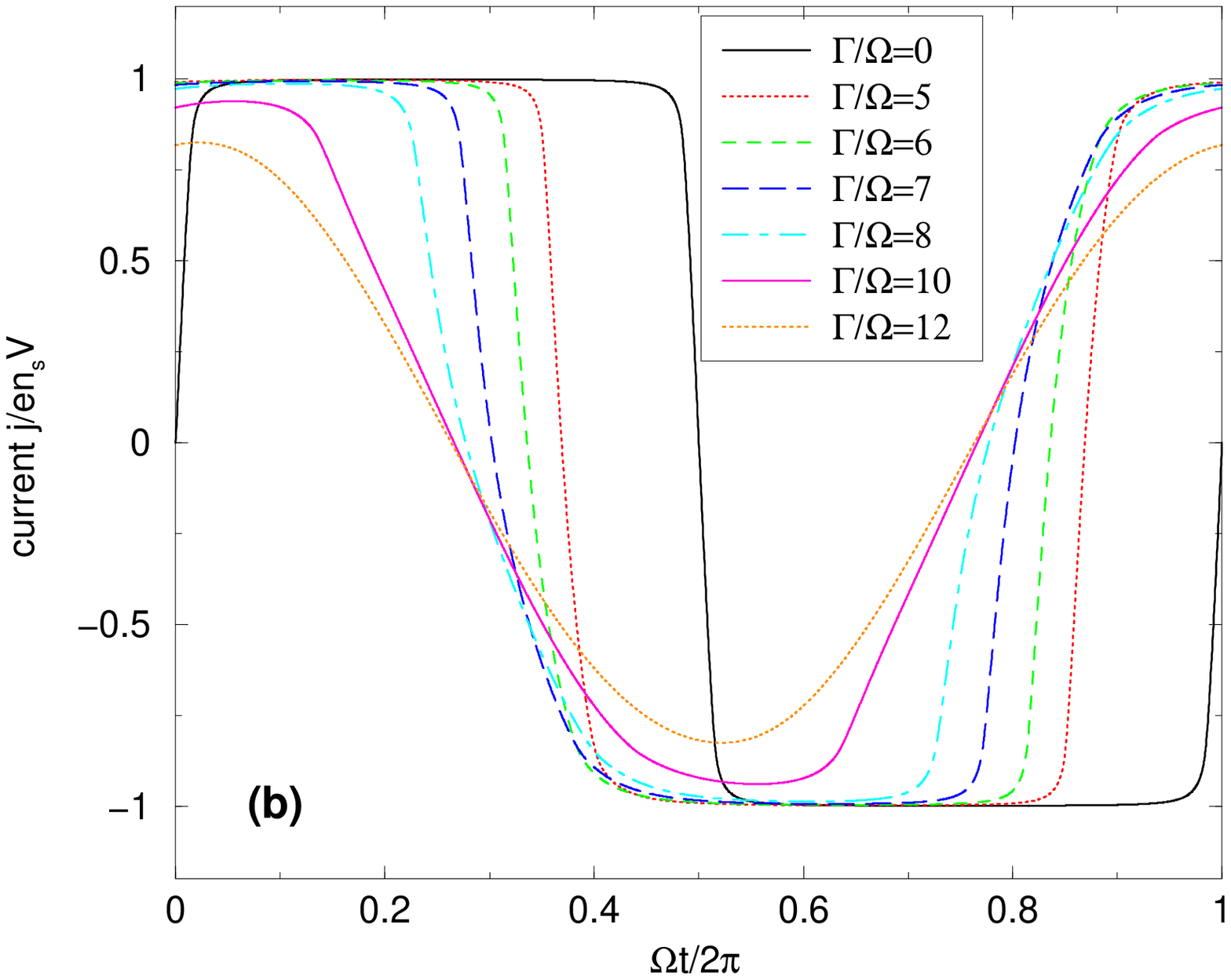}\\ 
\caption{\label{PJt} The time dependence of (a) the momentum $P(t)$ and (b) the current $j(t)/en_sV$ at the field parameter ${\cal E}=10$ at several values of the radiative decay $\Gamma/\Omega$. 
} 
\end{figure}

Now consider response of graphene to a harmonic excitation (\ref{harm}). 
It is convenient to rewrite Eqs. (\ref{eqp0a}) and (\ref{jx2}) in the form
\be 
\frac{d{\bf P}}{d\tau}+\frac{\Gamma}\Omega \frac{\bf P}{\sqrt{1+P^2}}{\cal G}(Q)  =\frac{ e{\bf E}_0V}{\mu\Omega}\cos \tau, \ \ \tau=\Omega t,\label{eqp0b} 
\ee 
\be
\frac{\bf j}{en_sV}=\frac{\bf P}{\sqrt{1+P^2}}{\cal G}(Q) ,\label{jb} 
\ee
from which one sees that the solutions depend on two dimensionless parameters, ${\cal E}=eE_0V/\mu\Omega$ and $\Gamma/\Omega$. If the field parameter ${\cal E}$ is small, the response is linear, $|P|\ll 1$, and  
\be
{\bf P}(t)=\frac {e{\bf E}_0}{p_F\sqrt{\Omega^2+\Gamma^2}}\cos\left(\Omega t-\arctan\frac\Omega\Gamma\right),\label{linsol} 
\ee 
\be
{\bf j}(t)=en_sV{\bf P}(t).
\ee
More interesting is the case of the strong electric fields, ${\cal E}\gg 1$. 
Figure \ref{PJt} shows the time-dependence of the momentum $P(t)$ and of the dimensionless electric current $j(t)/en_sV$ at ${\cal E}=10$ and several values of the radiative decay parameter $\Gamma/\Omega$. If $\Gamma/\Omega$ does not exceed the value of about ${\cal E}/2$ ($=5$ in the example of Fig. \ref{PJt}), the self-consistent-field effects lead only to the phase shift of the current, not influencing the shape of the current-time curves and hence not reducing the amplitudes of the higher harmonics. At higher values of $\Gamma/\Omega$ (between $\Gamma/\Omega\sim {\cal E}/2$ and $\Gamma/\Omega\sim {\cal E}$) the shape of the current-time curves smoothly modifies from the step-like form to the sinusoidal form, and at $\Gamma/\Omega\gtrsim {\cal E}$ the higher harmonics are fully suppressed. Quantitatively this can be seen from analysis of the Fourier harmonics of the current. Expanding the current in the Fourier series 
\begin{figure} 
\includegraphics[width=8.5cm]{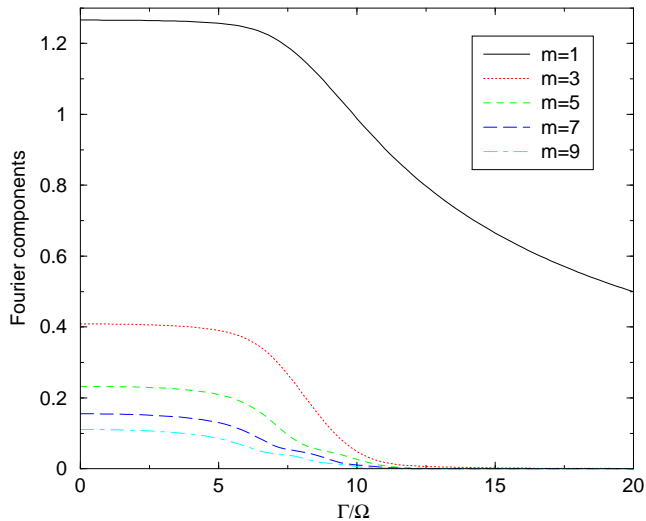}\\ 
\caption{\label{F10} Fourier harmonics of the current $j(t)/en_sV$ at ${\cal E}=10$ as a function of $\Gamma/\Omega$. 
} 
\end{figure} 
\be
\frac{j(t)}{en_sV}=\sum_{m}\left[A_m\cos(m\Omega t)+B_m\sin(m\Omega t)\right],
\ee
we calculate the amplitudes $F_m=\sqrt{A_m^2+B_m^2}$ and plot them as a function of $\Gamma/\Omega$ in Figure \ref{F10} for the lowest harmonics numbers from 1 to 9. One sees that, indeed, the higher harmonics are suppressed at $\Gamma/\Omega\gtrsim {\cal E}$, but almost not influenced by the radiative decay if $\Gamma/\Omega\lesssim 0.7 {\cal E}$. We have performed similar calculations at other values of the field parameter ${\cal E}$ and found the same results if ${\cal E}\gg 1$.

The condition of the effective frequency upconversion, with account of the self-consistent-field effects and the radiative decay, thus assumes the form ${\cal E}\gtrsim\Gamma/\Omega$, or
\be
E_0\gtrsim\frac{\mu\Gamma}{eV}=\frac{2\pi n_seV}c\approx 300\frac{\rm V}{\rm cm}\times n_s(10^{11}/{\rm cm}^2).\label{estim33}
\ee
This condition does not depend on the frequency of radiation.

It should be noticed that in the above consideration the sample was assumed to be infinite. Quantitatively, this means that the lateral dimensions of the graphene layer $L$ should exceed the wavelength of radiation $\lambda$. At the frequency of 1 THz the above derived  formulas will thus be valid for samples with the dimensions larger than $\simeq 300$ $\mu$m. As the currently available samples are smaller ($\simeq 10-30$ $\mu$m), the above consideration overestimates the role of the radiative decay. If $L\lesssim \lambda$, the radiative decay rate $\Gamma$ in the above formulas should be replaced by $\Gamma(L/\lambda)^2$, Ref. \cite{Mikhailov96a}. The upconversion effect should then be observed at electric fields lower than those given by the estimate (\ref{estim33}).

\section{Response of graphene to a pulse excitation}\label{timedom}

Now consider the response of graphene to a pulse excitation ${\bf E}^{ext}(t)={\bf E}_{0}\tau_0\delta(t)$, where ${\bf E}_{0}$ and $\tau_0$ are the amplitude and the duration of the pulse. The pulse amplitude ${\bf E}_{0}$ can be arbitrarily strong. Eq. (\ref{eqp0a}) now assumes the form
\be
\frac{d{\bf P}(t)}{dt}+\Gamma \frac{\bf P}{\sqrt{1+P^2}}{\cal G}(Q) 
=\frac{e}{p_F}{\bf E}_0\tau_0\delta(t).\label{eqp0c} 
\ee
The current is determined again by Eq. (\ref{jx2}). The solution of Eq. (\ref{eqp0c}) can be written in the following implicit form
\be
\Gamma t=\int_{P(t)}^{P_0}\frac{\sqrt{1+P^2}}{P{\cal G}(Q)}dP,\ \ Q=\frac{2P}{1+P^2},\label{implicit}
\ee
where $P(t)$ is the projection of the vector ${\bf P}(t)$ on the direction of the external electric field and 
\be 
P_0\equiv P(0)=\frac{eE_{0}\tau_0}{p_F}.\label{bc} 
\ee 
The function ${\cal G}(Q)$ is known from Appendix \ref{funcG}.

\begin{figure} 
\includegraphics[width=8.5cm]{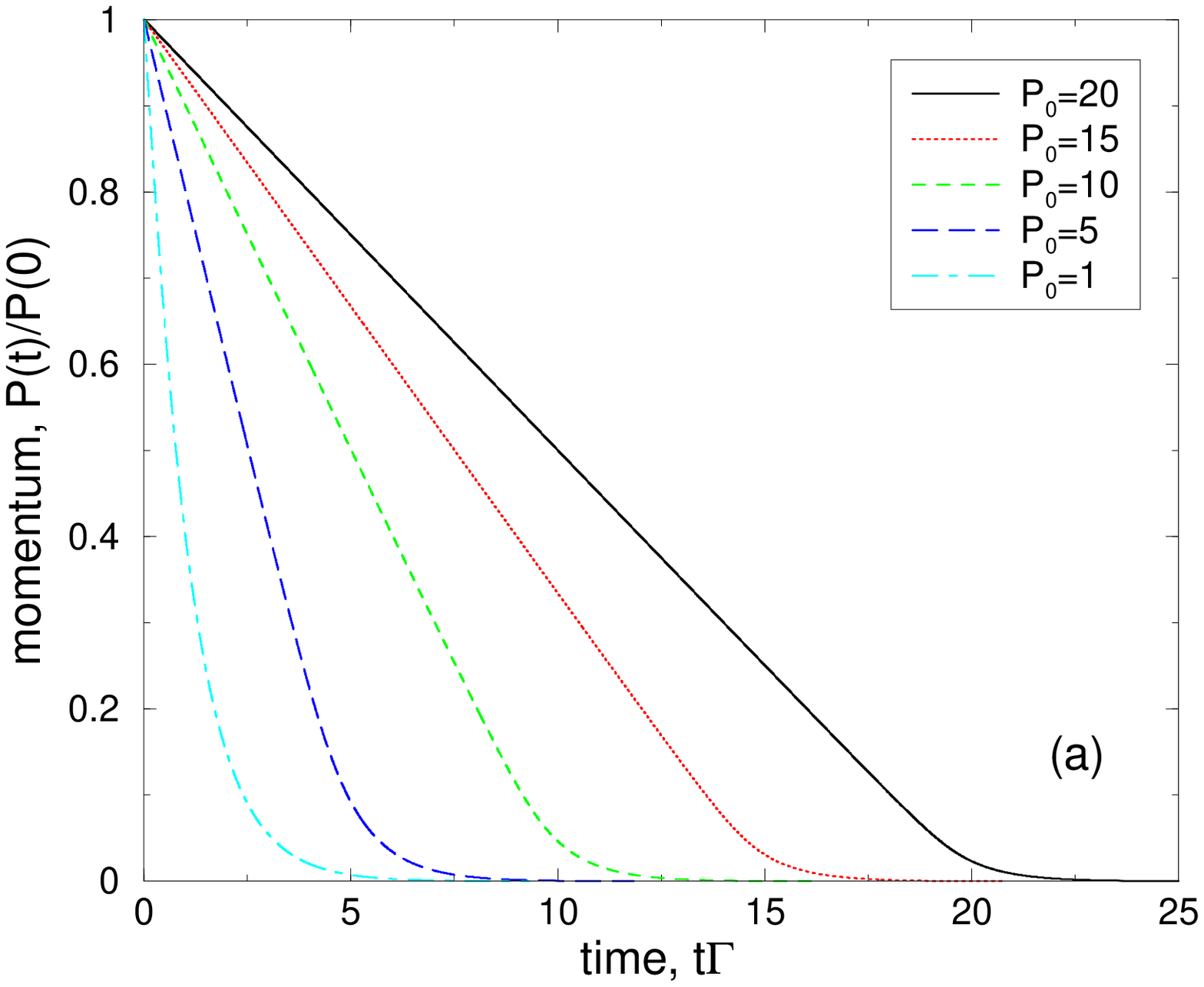}\includegraphics[width=8.5cm]{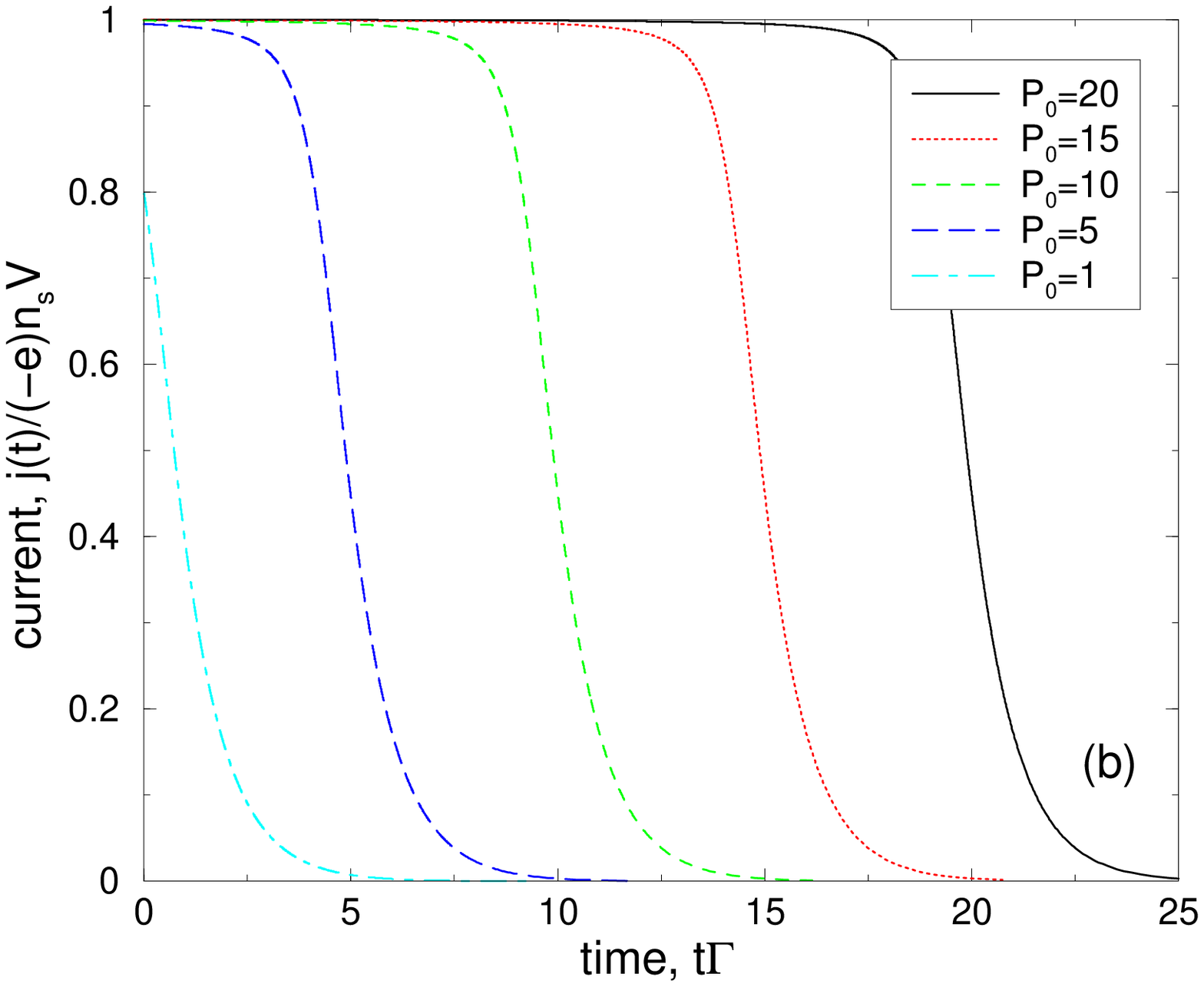} 
\caption{\label{fig1} (Color online) The time dependence of {\bf (a)} the momentum $P(t)/P_0$ and {\bf (b)} the electric current $j(t)/en_sV$, at a pulse excitation of graphene. Different curves correspond to different pulse amplitudes $P_0=eE_0\tau_0/p_F$. 
} 
\end{figure}

Figure \ref{fig1}a shows the dependence $P(t)$, calculated from Eq. (\ref{implicit}), at different values of the electric field parameter $P_0=eE_0\tau_0/p_F$. If the external field is small, $P_0\ll 1$, the system relaxes after the pulse excitation exponentially, similar to the conventional 2D electron systems with the parabolic dispersion, 
\be 
P(t)=P_0\exp(-\Gamma t), \ \ P_0\lesssim 1. \label{expresp} 
\ee 
The characteristic decay time is determined in this case by the inverse radiation decay rate (\ref{raddecgraph}). If the external field is strong, $P_0\gg 1$, the momentum of the system first decays linearly in time, 
\be 
P(t)=P_0-\Gamma t, \ \ P_0\gtrsim 1.\label{linresp} 
\ee 
The linear dependence (\ref{linresp}) remains valid until $P(t)$ reduces down to $P(t)\simeq 1$ (until $t\simeq P_0/\Gamma$); after that $P(t)$ decays exponentially like in (\ref{expresp}).  
The current ${\bf j}(t)$ in the strong excitation regime $P_0\gg1$ equals its highest possible value $en_sV$ and time-independent at $t\lesssim P_0/\Gamma$, and then exponentially decays (at the time scale $\simeq \Gamma^{-1}$) down to zero, as shown in Figure \ref{fig1}b.   

The fact, that after the pulse excitation electrons in graphene move with a constant velocity $V\approx 10^8$ cm/s for quite a long time $\sim P_0/\Gamma$, may have interesting applications. In a finite-size graphene sample such excited electrons will be reflected by the boundaries and oscillate in the sample with the typical frequency $\sim V/L$, lying in the terahertz range, if the sample dimensions $L\lesssim 1\ \mu$m. As at $P_0\gg 1$ the time $P_0/\Gamma$ is much longer than the oscillation period, this may lead to a coherent terahertz radiation from graphene excited by a strong pulse electric field.

%\newpage

\section{Summary and Conclusions}\label{concl}

We have developed the quasi-classical kinetic theory of the non-linear electromagnetic response of graphene. Electrodynamic equations, describing the self-consistent response of the system to a uniform time-dependent external electric field, have been derived and solved for the cases of harmonic and pulse excitation. The presented theory is valid at $\hbar\Omega\lesssim \max\{\mu,T\}$, which covers the frequency range up to $\sim 10-30$ THz for relevant experimental conditions. 

If the system is subjected to a pulse excitation and the amplitude of the external field is small (the linear regime), its response is described by a standard exponential decay with the characteristic decay rate $\Gamma$. The density dependence of the radiative decay rate in graphene ($\Gamma\propto n_s^{1/2}$) differs from that of conventional 2D electron systems ($\Gamma_{par}\propto n_s$). In the non-linear regime, when the pulse amplitude is strong, response of the system is not exponential. The average momentum of graphene electrons falls down linearly after the pulse excitation, with the average current remaining constant, during the time proportional to the amplitude of the external field and inversely proportional to $\Gamma$. 

The graphene layer excited by the harmonic electromagnetic wave with the frequency $\Omega$, re-emits radiation at higher harmonics $m\Omega$, $m=3,5,7,\dots$, with the efficiency falling down slowly, as $1/|m|$. The amplitude of the external electric field, required for getting into the non-linear regime, grows with the charge carrier density and is of order of several hundred V/cm for typical experimental parameters. The operating frequency of such a frequency multiplier can vary in a broad range, from microwaves up to mid-infrared.  The efficiency of the  frequency upconversion effect in graphene can be increased further by using the plasmon, the cyclotron, or the magnetoplasmon resonances. 

The predicted non-linear phenomena in graphene open up new exciting opportunities for building electronic and optoelectronic devices for terahertz and sub-terahertz part of the electromagnetic spectrum. 

\acknowledgments

This work was partly supported by the Swedish Research Council and INTAS. 

\appendix
\section{Function ${\cal G}(Q)$} \label{funcG}

The function 
\be
{\cal G}(Q)=\frac 4{\pi Q}\int_0^{\pi/2}\cos\theta d\theta\left(\sqrt{1+Q\cos\theta}-\sqrt{1-Q\cos\theta}\right)
\ee
can be expanded in powers of $Q$,
\be 
{\cal G}(Q)= \sum_{l=0}^\infty  
\frac{\Gamma(3/2)}{\Gamma(1/2-2l)}  
\frac{Q^{2l}}{2^{2l-1}l!(l+1)!}. 
\label{Gexpan}
\ee 
Since $|Q|=|2P/(1+P^2)|\le 1$, the expansion (\ref{Gexpan}) is valid at all values of $P$. The first terms of this expansion are
\be 
{\cal G}(Q)\approx 1+\frac 3{32}Q^2+\frac{35}{1024}Q^4\dots . 
\label{G} 
\ee 
If $Q=1$, then
\be 
{\cal G}(1)=\frac{8\sqrt{2}}{3\pi}\approx 1.200. 
\ee 

\bibliography{../../../BIB-FILES/mikhailov,../../../BIB-FILES/lowD,../../../BIB-FILES/thz,../../../BIB-FILES/graphene,../../../BIB-FILES/emp}

\end{document}